\documentclass[useAMS,usenatbib,usegraphicx]{mn2e}

\usepackage{epsfig}

\title[K-band transit and secondary eclipse photometry of exoplanet OGLE-TR-113b]{K-band transit and secondary eclipse photometry of exoplanet OGLE-TR-113b}
\author[Snellen \& Covino]{I.A.G. Snellen$^{1}$\thanks{E-mail:
snellen@strw.leidenuniv.nl}, \& E. Covino$^{2}$\\
$^{1}$Leiden Observatory, Postbus 9513, 2300 RA, Leiden, The Netherlands\\
$^{2}$ I.N.A.F., Osservatorio Astronomico di Capodimonte, Salita Moiariello 16, 80131 Napoli, Italy}
\begin{document}

\date{}


\maketitle


\begin{abstract}
We present high precision K-band photometry of the transit and secondary eclipse of 
extrasolar planet OGLE-TR-113, using the SOFI near-infrared instrument on ESO's NTT.
Data were taken in 5 second exposures over two periods of 3$-$4 hours, using random
jitter position offsets. In this way, a relative photometric precision of $\sim$1\% per frame was 
achieved, avoiding systematic effects that seem to become dominant at precisions exceeding this 
level, and resulting in an overall accuracy of 0.1\% per $\sim$10 minutes. 
The observations of the transit show a flat bottom light-curve indicative of a significantly lower
stellar limb-darkening at near-infrared than at optical wavelengths. The observations of the secondary eclipse result in a 
3$\sigma$ detection of emission from the exoplanet at 0.17$\pm$0.05\%. However, residual systematic errors 
make this detection rather tentative.
\end{abstract}


\section{Introduction}

While searching for extra-solar planets using the transit method seems less efficient than using the radial velocity technique, the scientific value of studying transiting planets has proven to be enormous. 
Firstly, the occurence of transits implies an almost edge-on orientation of the planetary orbit, allowing (in combination with radial velocity measurements) the determination of the planet's mass, radius and density (e.g. Charbonneau et al 2000; Konacki et al. 2003). Secondly, detailed follow-up observations of those planets transiting bright, nearby stars have lead to several scientific break-throughs. 
Atmospheric transmission spectroscopy of HD209458b using the Hubble Space Telescope have revealed the presence of 
a Sodium-rich  atmosphere (Charbonneau et al. 2002), and an evaporating exosphere containing Hydrogen, Oxygen, and Carbon 
(Vidal-Madjar et al. 2003; 2004). Furthermore, infrared Spitzer observations have recently resulted in the first measurements of direct thermal emissions from HD209458b, TrES-1, and HD189733b, by detection of their secondary eclipse (Deming et al. 2005; Charbonneau et al. 2005; Deming et al. 2006). 

Although {\sl near-infrared} (NIR) observations of transiting planets can also be of great scientific value, the possibilities in this wavelength regime have not yet been fully exploited (Richardson et al. 2003; 2006; Snellen 2005). 
Stellar limb darkening is significantly less pronounced at longer wavelengths, and therefore near-infrared transit photometry can 
improve the determination of the inclination of the orbit and subsequently the planetary radius. 
Also, in this way, NIR transit photometry can help to discriminate against false interlopers in transit surveys caused by stellar blends or grazing eclipsing binaries. 
Arguably most interestingly, the K-band is also open to secondary eclipse measurements. Although the planet/star flux ratio is less
favourable at this wavelength compared to the mid-infrared, the spectrum of a planet at 2.2 micron will be significantly
brighter than inferred from its black body temperature, such that  a hot Jupiter transiting a solar type star is expected to produce a secondary
eclipse in K-band with a depth of $\sim$0.1\% (e.g. Charbonneau et al. 2005).  Such measurements would form a vital part of the near-to-mid infrared spectrum,
required to reach detailed conclusions on the planets' effective temperatures, Bond albedos, and atmospheric physics.

For either transit or secondary eclipse K-band observations to be valuable, a photometric
precision in the order of 0.1\% or better is required. This is far from straightforward, since, using standard techniques, most 
near-infrared instruments typically only  show a photometric reliability down to the $\sim$1\% level, at
which systematic uncertainties, such as flat fielding errors, intra-pixel sensitivity variations, 
and/or background variations become dominant over the photon noise. 
First attempts by Snellen (2005) to reach a significantly better photometric precision at K involved 
observations of the secondary eclipse of HD209458b using the UK InfraRed Telescope (UKIRT). 
The telescope was defocused so that the starlight was spread out over a large part of the array. 
This not only avoided saturation of the K=6.3 star, but also diminished the systematic uncertainties in 
the photometry. The drawback of these observations were that no nearby, similarly bright comparison stars were available to perform relative photometry. Therefore observing cycles between the target 
and reference stars were needed, limiting the accuracy due to atmospheric absorption and
point-spread-function/seeing variations at the time-scale of the cycle time.
Although the secondary eclipse was not detected, it was shown that a photometric precision of 0.1\% is possible.

\subsection{The {\sl very} hot Jupiter OGLE-TR-113}

In this paper we present results on high precision K-band photometry, using the SOFI near-infrared camera on the European Southern Observatory's (ESO) New Technology Telescope (NTT), targeting the transit and secondary eclipse of the very hot Jupiter OGLE-TR-113.
OGLE-TR-113 was first identified as a candidate transiting extrasolar planet by Udalski et al. (2002)
within the OGLE transit survey (Optical Gravitational Lens Experiment), showing $\sim$2\% 
photometric dips at regular 1.4325 day intervals. Subsequently, Bouchy et al. (2004) and Konacki et al. (2004) have detected the 
accompanying radial velocity variations induced by the secondary, confirming it to be a 'very hot Jupiter' with a mass of 
$\sim$ 1.3 M$_{\rm{Jup}}$. The most recent ephemeris of OGLE-TR-113, based on more than 520 cycles with OGLE data\footnote{see http://www.astrouw.edu.pl/$\sim$ogle}, has now been determined to be
\[
\rm{HJD}_{\rm{min}}=2452325.79823+1.4324758\ E
\]
This is in close agreement with the ephemeris derived by Konacki et al (2004) and Gillon et al. (2006).

Although OGLE-TR-113  (Ks=12.86) is more than 6 magnitudes fainter than HD209458, it is sufficiently bright such that photon statistics allow a photometric 
precision  of 0.1\%  within 5 minutes. More importantly, it is located in the densely populated galactic plane, and several bright stars (Ks=11$-$12), 
that are required for calibration, are within 1$'$, allowing more accurate photometry than obtained for HD209458.
In sections 2 and 3 of this paper the observation strategy, and the data reduction and analysis are described. 
In sections 4 and 5 the results are presented and discussed.

\section{Observing strategy}
\begin{figure}
\hspace{-0.5cm} \psfig{figure=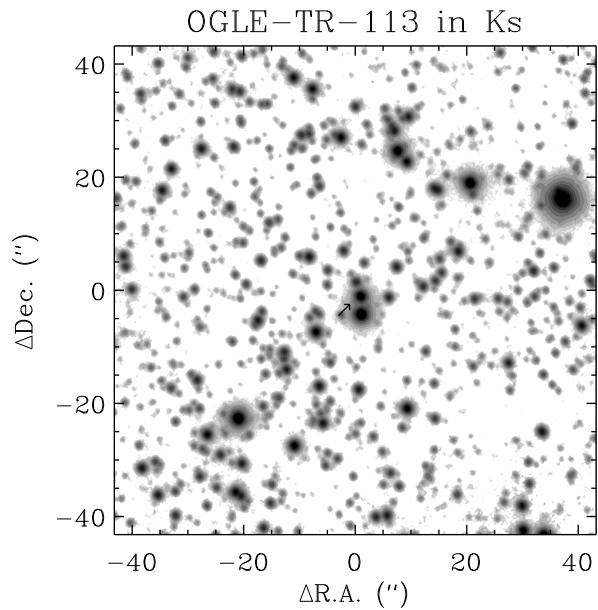,width=9cm}
\caption{\label{image} Central part of the image stack of the 1337 images taken on the night of March 15 2006. OGLE-TR-113 is indicated by the arrow. The main stars used for relative photometry are located
3$''$ S, 30$''$ NW, and 30$''$ NE from the target.  }
\end{figure}
\begin{figure}
\psfig{figure=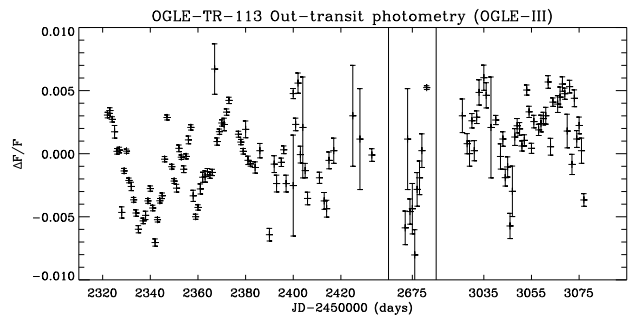, width=8.5cm}
\caption{Daily out-transit averages of the relative I-band flux of OGLE-TR-113, derived from data
provided by the OGLE team (Udalski et al. 2002). Flux variability at a level of 0.5\% is clearly visible. The associated time scale of $\sim$30 days makes it rather unlikely to be associated with systematic effects in the observations, and 
suggest it is caused by intrinsic variability of the host star. This I-band variability is removed from the data prior to the analysis performed in this paper. \label{out_transit}}
\end{figure}
Near-infrared observations are prone to time-dependent systematic effects in flat fielding, 
illumination corrections, intra-pixel sensitivities,  and possibly other not well understood effects.
This generally limits the photometric precision to a level of $\sim$1\%,  even if the photon statistics would 
allow a significantly higher accuracy. Since we require a precision of $<$0.1\% to be able
to obtain scientifically meaningful results, we used the following observing strategy. 
The exposure time for each frame was chosen so that purely the photon noise statistics for
OGLE-TR-113 would result in a relative photometric accuracy of $\sim$0.5\%, and such that the 
photometric precision per frame is dominated by systematic uncertainties. 
In addition, we used a random position jitter pattern to subsequently randomize these systematic errors. 
In this way, the photometric precision improves with the square-root of the number of frames,
and by keeping the exposure time short (short enough to keep the accuracy dominated by
systematic effects), a maximum precision per time step is achieved. 

We used the SOFI infrared camera on the NTT for 
our observations. The instrument was used in large field imaging mode, with the 0.288 arcsecond 
per pixel plate scale. It has a gain of 5.4 e/ADU, and 
a read out noise level of $\sim$2.1 ADU using the double correlated read out mode. All observations were performed in the Ks filter with exposure times
of 5 seconds (2 DIT $\times$ 5 second in the case of the transit). This filter has a central wavelength of 
2.162 $\mu$m and a FWHM of 0.275 $\mu$m. Data were taken on the nights 
of March 15 and 17, 2006, around a secondary eclipse and transit of OGLE-TR-113 respectively. Both nights were photometric, with a median seeing of 0.8 arcseconds.

To minimise the overheads of each exposure to 5$-$6 seconds, 
the array was windowed to 264$\times$264 pixels.
Each jitter position was chosen randomly within a 200$\times$200 pixel
box around the centre of the array, using an automated telescope control procedure. 
In this way $\sim 200-300$ images per hour were taken.
Assuming a circular planetary orbit, the scondary eclipse of OGLE-TR-113 was expected
at 13:22 UT on the night of March 15, 2006, and the target was observed from 11:50 UT until 
16:10 UT  (with a few small gaps caused by telescope software glitches). The transit occurred at 
16:55 UT on March 17, 2006, and was observed continuously from 15:30 UT until 18:20 UT.

\section{Data reduction and analysis}
\begin{figure}
\psfig{figure=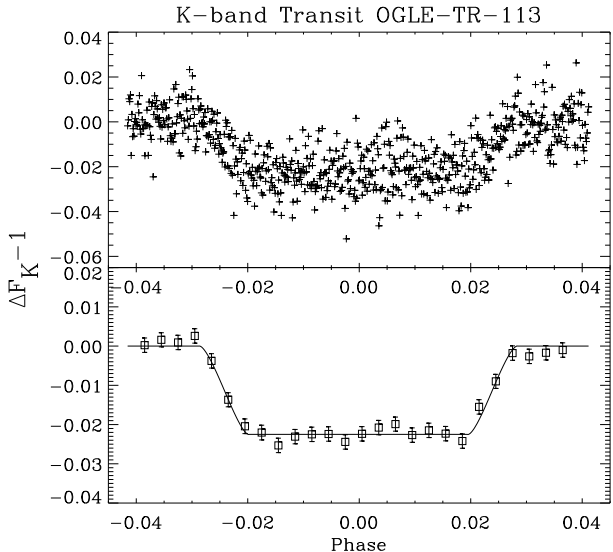, width=8.5cm}
\caption{K-band photometry of a transit of OGLE-TR-113. In the upper panel the 665 independent flux measurements are plotted, which show a scatter of 0.9\% around the best fitting model. The lower panel shows the same data , binned to 6 minutes (25 data points), resulting a dispersion of 0.16\%. Note how flat-bottomed the transit light curve is. This is the result of a low limb-darkening coefficient in K-band.\label{transit}}
\end{figure}
\begin{table}
\caption{Results of the least-square fit of the transit data, with the fitted parameters in column 1, 
and in column 2, 3, and 4, the fitted parameters and their associated 1$\sigma$ uncertainties 
for the K-band, I-band, and combined K+I band data.\label{results}}
\begin{tabular}{crrr}
Parameter & K-band & I-band & I+K band\\ \hline
\\
impact                         &0.5$^{+0.1}_{-0.1}$&0.1$^{+0.2}_{-0.1}$&0.4$^{+0.1}_{-0.2}$\\
parameter\\
\\
limb darkening             &0.00$^{+0.10}_{-0.00}$&0.55$^{+0.10}_{-0.10}$&K: 0.00$^{+0.20}_{-0.00}$\\
coefficient & & & I: 0.45$^{+0.10}_{-0.10}$\\
\\
R$_{\rm{p}}$/R$_{\rm{*}}$&0.150$^{+0.001}_{-0.002}$&0.147$^{+0.002}_{-0.003}$&0.151$^{+0.002}_{-0.002}$\\
\\
R$_*$M$_{*}^{-1/3}$                    &0.93$^{+0.04}_{-0.04}$ & 0.85$^{+0.03}_{-0.03}$ & 0.89$^{+0.03}_{-0.03}$\\
\\ \hline
\end{tabular}
\end{table}
The observations of March 15 resulted in 1337 images, which were reduced in the following way. 
First for each image a sky map was produced by averaging the 10 images taken nearest in time, after excluding those pixels that deviate by more than 2$\sigma$ from the pixel-mean. Other, more complicated procedures to optimally remove the stars from the sky-image were tried but did not improve the results.  These sky maps were then
subtracted from each image. A flat field was created by exposing alternately an illuminated and 
unilluminated dome panel, taking the difference of the two, and normalising it to 1.
Subsequently, the sky subtracted images were divided through this flat field. 
The observations of March 17, that resulted in 665 images, were reduced in a similar manner.
The central part of the image stack of the March 15 observations is shown in figure \ref{image}.

Aperture photometry was performed on the target star and a dozen other stars in the direct 
vicinity, using a 3 pixel radius. Note that more 
complicated psf-fitting procedures were not an improvement, due to PSF variations in time and 
across the array. For each image the positions of the 
stars were fitted independently, resulting in relative positional accuracies of the stars
of $\sim$0.03 pixel, small enough not to significantly influence the photometric measurements.

The flux of the target star reached in each frame a count level of approximately 
20,000-25,000 ADU, depending on the seeing. The background level was typically 3000 ADU/pixel.
With a gain of g=5.4 e$^-$/ADU, this means photometric uncertainties of 0.7\% per image (0.5\% for the March 17 observations) are expected from the noise statistics. Three bright stars directly in the vicinity of 
the target are useful for relative photometry, with count levels between 40,000 and 60,000.
Peak levels were found not to exceed 10,000-12,000 ADU and possible non-linearities of 
the array were not found to be a problem. One of the three reference stars (the only 
one that is always on the array) is located 3.2 arcseconds south from the target. It was found that 
the outer wings of the point spread function were influencing the target photometry at the 
0.3$-$0.5\% level, dependent on the seeing, and the rotation angle of the detector (the NTT is 
alt-az mounted). The latter is due to the four radial spikes in the PSF caused by the 
secondary mirror support. These dependencies were measured and taken out by determining
the flux within a 3 pixel aperture, 3.2 arcseconds north from a bright 200,000 ADU/frame star 
throughout the night (40$''$ W of the target as shown in figure \ref{image}). 
A similar technique (Covino et al. 2004), that measures the sky  level on the diametrically 
opposite position of the star involved, in this case works less well, because there appears to be an asymmetric component to the wings of the PSF, and since at the count levels the observations are sky noise limited.

To be able to make a full comparison of our K-band photometry with optical results, 
the I-band photometry obtained by the OGLE team containing a total of 1517 measurements covering 14 transits 
(Udalski et al 2002; http://www.astrouw.edu.pl/$\sim$ogle), were first reanalysed.
We found that the scatter in the light curve (excluding those points within the transits) was significantly
greater than the uncertainties quoted by the OGLE team. This appears to be caused by intrinsic variability of the star. In figure \ref{out_transit} the nightly averaged  fluxes  of OGLE-TR-113 are shown,
including only those flux points at phase, $|\phi|>0.04$. Clear variations are visible with an amplitude of $\sim$0.005, on a time scale of $\sim$30 days, and suggests that it is linked with the rotation of the host star, and it makes it rather unlikely to be an observational systematic effect. This variability was removed from the data prior to the transit fitting described in the next section, by subtracting the daily out-transit averages from the 
flux points.

\section{Results and discussion}

The K-band observations of the transit have resulted in 665 independent flux measurements taken
within a period of 170 minutes (one every $\sim$15 seconds). The dispersion around the best fitting model (see below) is 0.9\%. The observations of the secondary eclipse resulted in 1337 measurements taken within 260 minutes (one per $\sim$11 seconds). These show a dispersion of 1.1\% around the mean.
The photometric precision achieved on both nights is consistent with a systematic uncertainty of 0.8\% per frame, added in quadrature to the photon noise. 

\subsection{K-band photometry of the transit}

The results of the K-band photometry of the transit of OGLE-TR-113 are shown in figure \ref{transit}. 
The model is derived with least square fitting using the equations of Mandel \& Agol (2002), with as free parameters the timing of the transit, $t_0$, one linear limb-darkening coefficient, $\pi$, the impact parameter of the transit, $i$, the planet/star size-ratio, $r$, and R$_*$M$_{*}^{-1/3}$,
where M$_*$ and R$_*$ are the stellar mass and radius. The last parameter links the orbital period of the planet with its orbital radius, and the duration of the transit. In this way no parameters are kept fixed.
The centre of the transit is found to be within 1$\sigma$ from that expected from the most up to date
ephemeris of OGLE-TR-113. The fitted parameters with their associated errors are given in column 2 of table \ref{results}. Both M$_*$ and R$_*$ are in units of solar mass and radius.  

 Most striking is that the bottom of the K-band light curve appears remarkably flat.
We investigated whether this is consistent with current models
of near-infrared limb-darkening. For this we used the K-band limb-darkening
parameters from Claret 2000, assuming a star with a surface gravity of g=4.0,
a temperature T=4750 K, and a metallicity of log [Fe/H]=0.1 (this model
produces light curves like that for a linear limb darkening model with
$\pi$=0.3 to within a fraction of a milli-magnitude). The best least-squared
fit to this model (varying again the impact parameter and planet/star size
ratio), results in a $\Delta \chi^2$=2.7 higher than that for the best
fitting linear limb darkening model (with $\pi$=0.0). It indicates that
this Claret model can be rejected at a 90\% confidence level. More
near-infrared transit light curves will need to be taken to establish
whether this is a valid conclusion.

\subsection{Comparison with optical light curve}

\begin{figure}
\psfig{figure=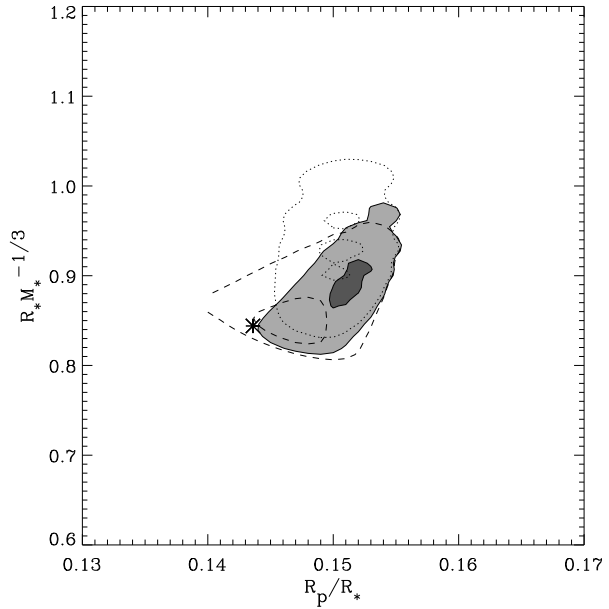, width=8.5cm}
\caption{\label{rm} The 67\% and 95\% confidence intervals for the fitted planet/star size ratio R$_{\rm{p}}$/R$_*$ versus R$_*$M$_*^{-1/3}$, from the K-band photometry (dotted lines), OGLE I-band photometry (dashed line), and the combined photometry (solid lines and shaded). The asterisk indicates the parameters as derived by Konacki et al. (2004).}
\end{figure}

Interesting is the comparison between the near-infrared and optical transit light curves.
Table \ref{results} gives the least-squares fitted parameters for the K-band data, the I band data from the OGLE team, and the combined K and I band data. It shows that all fitted parameters are within 1$-$2$\sigma$ 
consistent between K and I band. Figure \ref{rm} shows the confidence intervals for the planet/star size ratio
and R$_*$M$_{*}^{-1/3}$, for the K-band, I band, and combined data sets. The fit to the K-band photometry gives
a slightly larger size ratio and stellar parameters. This could be caused by a small flux contribution in I-band 
from the nearby bright star, which is carefully removed in K (where it is also of less influence due to the 
superior seeing). This could also be the reason why the K-band data are fit better with a larger impact 
parameter ($i$=0.5$\pm$0.1) than the I-band data ($i$$<$0.3).
Note however that due to the small planet to star
size ratio, an inadequate treatment of the stellar limb-darkening
(and/or non-spherical shape of the star) may also introduce systematic
effects in the determination of the derived parameters (in particular
of the impact parameter).

\begin{figure}
\psfig{figure=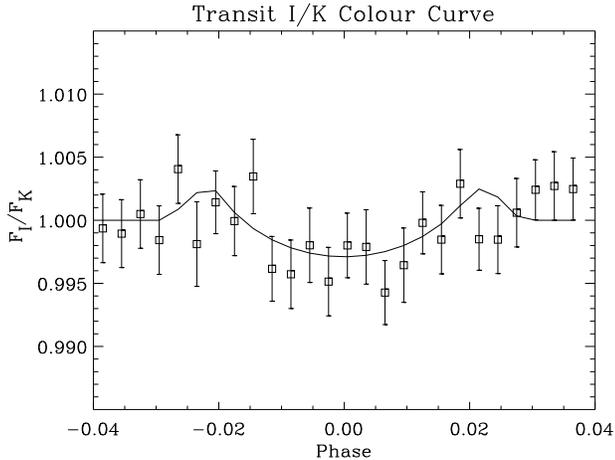, width=8.5cm}
\caption{\label{colour} The I/K band flux ratio during the transit. The fluctuations in colour are the result of the difference in limb darkening coefficient between I and K band.}
\end{figure}

As discussed above, one of the interesting features of the K-band light curve is the flat bottom of the transit, indicative of a low 
limb-darkening effect at this wavelength, confirmed by the least-squares fit to the data. Its contrast to the 
optical light curve is most visible by plotting the I/K flux ratio as function of orbital phase, as shown in figure \ref{colour}.
The fluctuations in this colour curve are indicative of the difference in limb-darkening coefficient between
I and K band, and the impact parameter of the transit (the solid line is the best fitting model). 
For example, if the planet passes the centre of the star, it will block out a larger fraction of the 
star light in the case of strong limb darkening than with weak limb darkening. However, near the stellar limb this effect will be the opposite. 

\subsection{K-band photometry of the secondary eclipse}

\begin{figure}
\psfig{figure=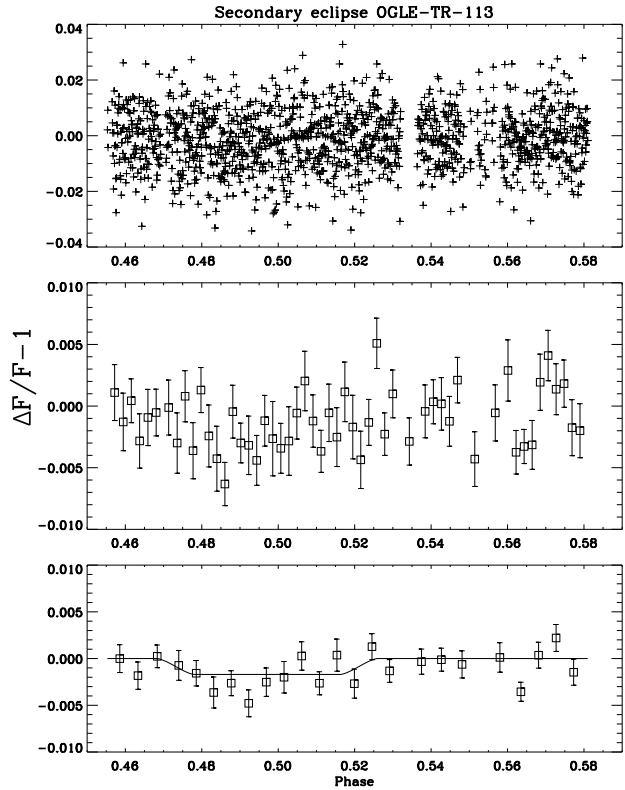, width=9cm}
\caption{\label{second} K-band photometry of the secondary eclipse. The upper, middle, and lower panels show the
unbinned, 5-minutes binned, and 10 minutes binned data respectively. The solid line shows the best fitting model 
with an amplitude of 0.17$\pm$0.05\% (2.8$\sigma$). the low statistical significance of the eclipse, and the presence of correlated noise on time scales of 5$-$10 minutes make this detection rather tentative.}
\end{figure}

The results from the secondary eclipse K-band photometry of OGLE-TR-113 are shown in figure \ref{second}.
In the upper panel the 1337 individual flux measurements are plotted, while in the middle and lower panel the same data are shown, but binned over 5 and 10 minutes (25 and 55 data points) respectively. The 5 and 10 minutes binned data show a scatter around the mean of 0.24\% and 0.17\% respectively. Since the individual data points show a dispersion of 1.1\%, this level about 10$-$20\% higher than expected, suggesting that there may be some correlated noise present on time scales of 5$-$10 minutes. This is indeed best visible in the middle panel of figure \ref{second}.

Although it is highly desirable to know the origin of this correlated noise 
(such that possibly its influence can be reduced in future observations), 
its source is unclear. The randomization of the jitter offsets 
has minimised the possible contributions caused by errors in flat fielding,
fluctuations in intra-pixel sensitivities, and dark current and PSF variations 
over the array. However, of course, time dependent effects related to the 
Earth's atmosphere are not reduced in this way. A colour dependence of 
telluric absorption and its time variability could influence the relative 
photometry. The aperture corrected source counts of bright isolated stars 
in the field vary by less than 4\%, indicating that the telluric absorption 
is very stable over the time scale of the observations and that any dependence 
of colour would be $<<$4\%. However, the relative flux density of OGLE-TR-113 
is found to be independent of the detected source counts from the bright stars 
in the field, making this an unlikely source for the correlated noise. 
Although variations in the seeing and their effect on the relative photometry 
have been corrected for, a small residual correlation
between the seeing and the relative photometry of the target remains.
However its effect is too small for being the source of the correlated noise. 
Errors in the sky subtraction could also be the source of the 
correlated noise, in particular because we have used a set of 10 images 
around each frame to determine the sky background. However also here, we 
believe that any residual error in sky subtraction can not result in the level 
of correlated noise as seen in the data, leaving its origin as a question mark.

\begin{figure}
\psfig{figure=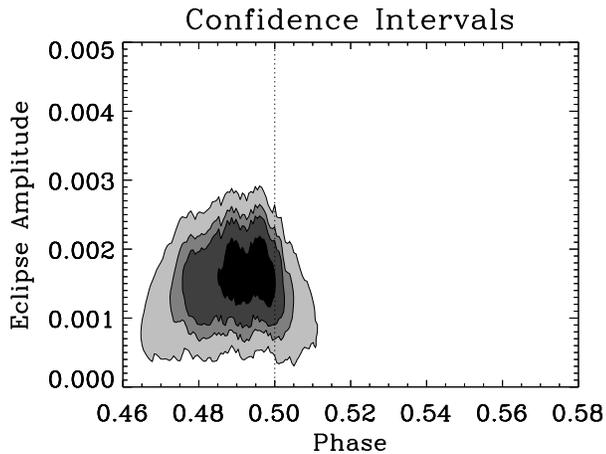, width=8.5cm}
\caption{\label{confi} The 68\%, 90\%, 95.4\% and 99.7\% confidence intervals for the amplitude and timing of the secondary eclipse in K-band. It shows that the timing of the eclipse is within 1$\sigma$ statistically in agreement with a circular orbit. }
\end{figure}

Since a possible dip can be seen in the light curve at a phase of 0.48-0.50, we fitted 
the data with an eclipse, with the timing and eclipse-depth as free parameters, but 
keeping the duration of the eclipse fixed (as derived from the transit data)
The secondary eclipse is statistically detected at a level of 2.8$\sigma$ at 
0.17\%$\pm$0.05\%. The central timing of the eclipse is determined at a 
phase between 0.485 and 0.500 (1$\sigma$), hence would be consistent with a 
circular planetary orbit. If the central phase is forced to 0.5 exactly, the secondary
eclipse signal is detected at 0.15\%$\pm$0.05\% (2.6 $\sigma$).
 However, the low statistical significance of the eclipse, and the presence of correlated noise on time scales of 5$-$10 minutes makes this detection rather tentative. The formal confidence intervals for the eclipse amplitude and timing are shown in figure \ref{confi}.

\section{Conclusions}
In this paper we present near-infrared Ks-band photometry of the transit and secondary eclipse of extrasolar planet OGLE-TR-113. We achieve a relative photometric precision of 0.1$-$0.2\% per 10 minutes of observations by taking many short exposures with random positional offsets.
The flat bottom of the transit light curve  shows that limb darkening is significantly less
pronounced in the near-infrared than in the optical, and that these observations help to constrain the 
impact parameter of the transit. Near-infrared observations like presented here would be adequate to distinguish between genuine transits and false detections caused by possible blends with eclipsing binary systems or grazing eclipses.  
The photometry of the secondary eclipse indicates a formal detection at a level of 0.17$\pm$0.05\% (2.8$\sigma$), and 0.15\%$\pm$0.05\% (2.6 $\sigma$) assuming a circular orbit. However, the low statistical significance and the presence of correlated noise on time scales of 5$-$10 minutes make this detection rather tentative.

\section*{Acknowledgments}
The authors wish to thank the ESO staff, in particular Valentin Ivanov,  for help and suggestions during the project. We thank the anonymous referee for 
useful comments and carefully reading the manuscript.
 The observations in this paper were collected at the European Southern Observatory, La Silla, Chile, using SOFI at the 3.5m NTT, within the observing program 076.C-0674(A).

\end{document}